\documentclass[fleqn,usenatbib]{mnras}

\usepackage{newtxtext,newtxmath}

\usepackage[T1]{fontenc}

\DeclareRobustCommand{\VAN}[3]{#2}
\let\VANthebibliography\thebibliography
\def\thebibliography{\DeclareRobustCommand{\VAN}[3]{##3}\VANthebibliography}

\usepackage{graphicx}	
\usepackage{amsmath}	

\title[Photo-$z$ estimation with LSTM Networks]{Photometric Redshift Estimation for CSST Survey with LSTM Neural Networks}

\author[Z. Luo et al.]{
Zhijian Luo$^{1}$\thanks{E-mail: zjluo@shnu.edu.cn},
Yicheng Li$^{1}$,
Junhao Lu$^{1}$,
Zhu Chen$^{1}$,
Liping Fu$^{1}$,
Shaohua Zhang$^{1}$,
Hubing Xiao$^{1}$,
\newauthor
Wei Du$^{1}$,
Yan Gong$^{2,3}$,
Chenggang Shu$^{1}$,
Wenwen Ma$^{1}$,
Xianmin Meng$^{2}$,
Xingchen Zhou$^{2}$
\newauthor
and Zuhui Fan$^{4}$
\\
$^{1}$Shanghai Key Lab for Astrophysics, Shanghai Normal University, Shanghai 200234, China\\
$^{2}$Key Laboratory of Space Astronomy and Technology, National Astronomical Observatories, Chinese Academy of Sciences,\\
20A Datun Road, Beĳing 100101, China\\
$^{3}$Science Center for China Space Station Telescope, National Astronomical Observatories, Chinese Academy of Sciences,\\ 
20A Datun Road, Beĳing 100101, China\\
$^{4}$South-Western Institute for Astronomy Research, Yunnan University, Kunming 650500, China\\
}

\date{Accepted XXX. Received YYY; in original form ZZZ}

\pubyear{2015}

\begin{document}
\label{firstpage}
\pagerange{\pageref{firstpage}--\pageref{lastpage}}
\maketitle

\begin{abstract}
Accurate estimation of photometric redshifts (photo-$z$s) is crucial for cosmological surveys. Various methods have been developed for this purpose, such as template fitting methods and machine learning techniques, each with its own applications, advantages, and limitations. In this study, we propose a new approach that utilizes a deep learning model based on Recurrent Neural Networks (RNN) with Long Short-Term Memory (LSTM) to predict photo-$z$. Unlike many existing machine learning models, our method requires only flux measurements from different observed filters as input. The model can automatically learn the complex relationships between the flux data across different wavelengths, eliminating the need for manually extracted or derived input features, thereby providing precise photo-$z$ estimates. The effectiveness of our proposed model is evaluated using simulated data from the Chinese Space Station Telescope (CSST) sourced from the Hubble Space Telescope Advanced Camera for Surveys (HST-ACS) and the COSMOS catalog, considering anticipated instrument effects of the future CSST. Results from experiments demonstrate that our LSTM model, compared to commonly used template fitting and machine learning approaches, requires minimal input parameters and achieves high precision in photo-$z$ estimation. For instance, when trained on the same dataset and provided only with photometric fluxes as input features, the proposed LSTM model yields one-third of the outliers $f_{out}$ observed with a Multi-Layer Perceptron Neural Network (MLP) model, while the normalized median absolute deviation $\rm \sigma_{NMAD}$ is only two-thirds that of the MLP model.  This study presents a novel approach to accurately estimate photo-$z$s of galaxies using photometric data from large-scale survey projects.

\end{abstract}




\section{Introduction}

Since the pioneering work of \citet{baum1962photoelectric} on utilizing multi-band photometric data for estimating galaxy redshifts, the concept of photometric redshift (photo-$z$) has gained widespread recognition and has been applied in many large astronomical surveys \citep{baum1962photoelectric,baldwin1982iau,koo1985optical,loh1986photometric,connolly1995slicing,benitez2000bayesian,massarotti2001new,mobasher2007photometric,carrasco2013tpz,salvato2019many, hernan2021minijpas,newman2022photometric,euclid2023selection}. For instance, in the Legacy Survey of Space and Time (LSST) conducted by the Vera C. Rubin Observatory, photo-$z$ methods are employed to process a massive amount of observational data, thereby improving the accuracy of galaxy redshift estimates. The Euclid satellite also plans to utilize photo-$z$s to explore dark energy and the structure of the universe. Additionally, the Hyper Suprime-Cam Subaru Strategic Program (HSC-SSP) has adopted photo-$z$ techniques in its data analysis to support its cosmological research. Over the past few decades, research on photo-$z$ has significantly advanced from being considered a supplemental method when spectroscopic redshift (spec-$z$) was not feasible, to becoming a primary tool for determining redshifts in modern cosmological surveys.

Compared to spec-$z$ samples, photo-$z$ samples generally have simpler and more consistent selection functions, encompassing fainter magnitude limits and wider angular scales. This characteristic allows for the exploration of larger cosmic volumes, facilitating more comprehensive and accurate cosmological investigations \citep{hildebrandt2010phat}. The scope of photo-$z$ samples typically exceeds that of comparable spectroscopic samples by several orders of magnitude.

Today, many major ongoing and planned large-scale sky survey projects utilize photo-$z$ techniques as a crucial method to achieve their scientific objectives alongside the use of spectroscopic surveys that also cover a significant portion of the sky. These projects include the Sloan Digital Sky Survey (SDSS) \citep{fukugita1996sloan,york2000sloan}, the Dark Energy Survey (DES) \citep{collaboration2016more,abbott2021dark}, the Kilo-Degree Survey (KiDS) \citep{de2013kilo}, the Legacy Survey of Space and Time (LSST) \citep{ivezic2019lsst,abell2009lsst}, the Euclid Space Telescope \citep{laureijs2011euclid}, the Wide Field Infrared Survey Telescope or Nancy Grace Roman Space Telescope (WFIRST) \citep{spergel2015widefield,green2012wide,akeson2019wide}, the Hyper Suprime Cam Subaru Strategic Program (HSC-SSP) \citep{aihara2018hyper}, and the Roman Space Telescope survey \citep{akeson2019wide}, among others.

Despite their significance in extragalactic and cosmological research, the precision and accuracy of photo-$z$s present challenges. Factors such as filter choice, data quality, observational corrections, and estimation methodologies impact the accuracy of photo-$z$s. Of these factors, the methodology for estimating photo-$z$s plays a crucial role, as emphasized in the Euclid photo-$z$ challenge (see \citealt{desprez2020euclid}).

To address these challenges, researchers have developed various methods for calculating photo-$z$s and continuously refine these methodologies. Broadly, two main techniques are used for determining galaxy photo-$z$s: the template fitting method and the training set method.

The template fitting method, also known as spectral energy distribution (SED) fitting, involves matching template galaxy SEDs to observational photometric data to derive photo-$z$s \citep{lanzetta1996star,fernandez1999new,bolzonella2000photometric}. This approach is grounded in theory and can predict physical characteristics such as stellar mass, star formation history, and dust extinction. Notable SED fitting methods include HyperZ \citep{bolzonella2011hyperz}, BPZ \citep{benitez2000bayesian}, ZEBRA \citep{feldmann2006zurich}, EAZY \citep{brammer2008eazy}, LePhare \citep{arnouts1999measuring}, and BCNZ2 \citep{eriksen2019pau}.

In contrast, the training set method, or machine learning approach, establishes empirical relationships between redshift and galaxy properties like magnitude, color, and morphology to derive photo-$z$s \citep{collister2004ANNz,sadeh2016annz2,hoyle2015anomaly}. This method requires a training sample with known redshift values and utilizes machine architectures to optimize processing of input photometric data for accurate redshift predictions. Common machine learning techniques include artificial neural networks (ANN) \citep{firth2003estimating,collister2004ANNz,sadeh2016annz2}, support vector machines (SVM)  \citep{wadadekar2004estimating}, self-organizing maps \citep{way2012can,geach2012unsupervised,carrasco2013tpz}, Gaussian process regression \citep{way2006novel}, genetic algorithms \citep{hogan2015gaz}, k-nearest neighbors algorithm (kNN) \citep{ball2007robust}, boosted decision trees \citep{gerdes2010arborz}, random forest (RF) \citep{carrasco2014som,rau2015accurate,mucesh2021machine,lu2024estimating}, and sparse Gaussian frameworks \citep{almosallam2016sparse}. Additionally, deep learning methods like multi-layer perceptrons (MLP) \citep{zhou2021spectroscopic}, convolutional neural networks (CNN) \citep{hoyle2016measuring,d2018photometric,pasquet2019photometric}, and Bayesian neural networks (BNN) \citep{zhou2022photometricBNN} have found application in photo-$z$ estimation.

In practical applications, SED fitting methods may struggle to provide precise redshift estimates due to various factors such as template selection, corrections for extinction and reddening, as well as other potential sources of error. On the other hand, machine learning methods bypass the need to account for these complex factors. Given a sufficient number of unbiased training samples, machine learning techniques are capable of providing highly precise redshift estimates. Consequently, the utilization of machine learning approaches has become increasingly common in modern large-scale astronomical survey projects.

For machine learning models, selecting input features is crucial, especially when the input data is sequential. Researchers often need engage in feature engineering to determine which photometric features or derived features will provide optimal redshift prediction capability, subsequently utilizing these features as inputs for the model. However, these processes are highly intricate and challenging.

An alternative approach that avoids feature selection is using a Convolutional Neural Network (CNN) model \citep{hoyle2016measuring, zhou2021spectroscopic, zhou2022photometricBNN}. This model passes images of galaxies in all bands through the network without the need to select specific input features or even conduct photometric measurements. However, deep neural networks containing CNNs necessitate computing resources significantly greater than other traditional machine learning models, limiting their usage to predicting smaller-scale datasets.

In most cases, when using machine learning models for predicting photo-$z$s, researchers often use photometric fluxes from various bands as inputs without intricate feature engineering. Nonetheless, studies indicate that solely using photometric fluxes typically leads to inaccurate photo-$z$ predictions, necessitating the incorporation of features like colors to enhance accuracy \citep{lu2024estimating, zhou2021spectroscopic, fotopoulou2018cpz}. The primary reason for this is that colors can more directly reflect the shape and breakpoint features of a galaxy’s spectral energy distribution (SED) (e.g., Balmer/4000A break, Lyman break), providing more direct information related to redshift. This enhances the model’s ability to estimate the redshift of celestial objects with greater accuracy and reliability.

It is worth noting that colors are not directly observed but derived from photometric fluxes. Essentially, colors can be derived from the fluxes in different bands and do not provide additional information beyond what is already contained in the photometric fluxes. Therefore, traditional machine learning models need to incorporate colors as input parameters, indicating that these model structures cannot capture the correlations between input photometric data, necessitating the manual combination of photometric fluxes into color features for model inputs.

However, colors alone may not completely capture the correlations between fluxes measured in different bands. Moreover, determining which colors can be effectively used as inputs for the model remains challenging and is contingent upon factors such as the redshift range of the study samples and the specific characteristics of the observed filters. For instance, \citet{lu2024estimating} conducted a thorough analysis of the “input feature importance” for photo-$z$ estimation using simulated data from the Chinese Space Station Telescope (CSST). This study underscored the critical significance of selecting the most relevant color features and implementing effective processing methods.

In this study, we have proposed a new model based on Recurrent Neural Network (RNN) with Long Short-Term Memory (LSTM) for estimating photometric redshifts. The proposed model avoids the input feature selection issues commonly faced by most traditional machine learning models. It only requires organizing the photometric fluxes from each band in wavelength order to form an input sequence. The model can automatically and effectively learn the dependencies between them, capturing patterns and correlations within the flux sequences to achieve highly accurate predictions of photometric redshifts. 

Due to the inherent sequence processing capabilities of the LSTM model, it can effectively capture the correlations between the various data points in a sequence, thereby reducing the need for feature engineering. As a result, the proposed model no longer requires manually combining input features to generate colors and the like, nor does it require complex feature selection engineering, simplifying the feature selection process and avoiding human errors or subjective bias. Furthermore, compared to CNNs, our LSTM model does not need to process large image data, which leads to lower computational resource requirements and faster training speeds. Specifically, we conducted a series of experiments on our LSTM model using a computer equipped with an NVIDIA RTX 3090 GPU. With a batch size set to 128, the time taken per epoch for the LSTM model was approximately 1 second, and the total training time was about 17 minutes. We applied this method to simulated data from the Chinese Space Station Telescope (CSST) and compared it with other photometric redshift estimation methods.

The structure of this paper is as follows: In Section 2, a brief overview of the CSST survey is provided, along with an explanation of the process of generating the mock photometry catalog used in this study. Section 3 elaborates on the architecture and training process of the adopted LSTM neural network model. This section primarily features technical content related to the LSTM neural network. Readers who are not interested in the specific implementation details of the LSTM or who wish to focus directly on the research results may skip to Section 4. Section 4 presents the results of this model in estimating photo-$z$s on CSST simulated data and compares them with other models. Finally, Section 5 summarizes our research findings and conducts discussions.

\section{mock data}\label{section:mock}

\begin{figure}
	\includegraphics[width=1.0\columnwidth]{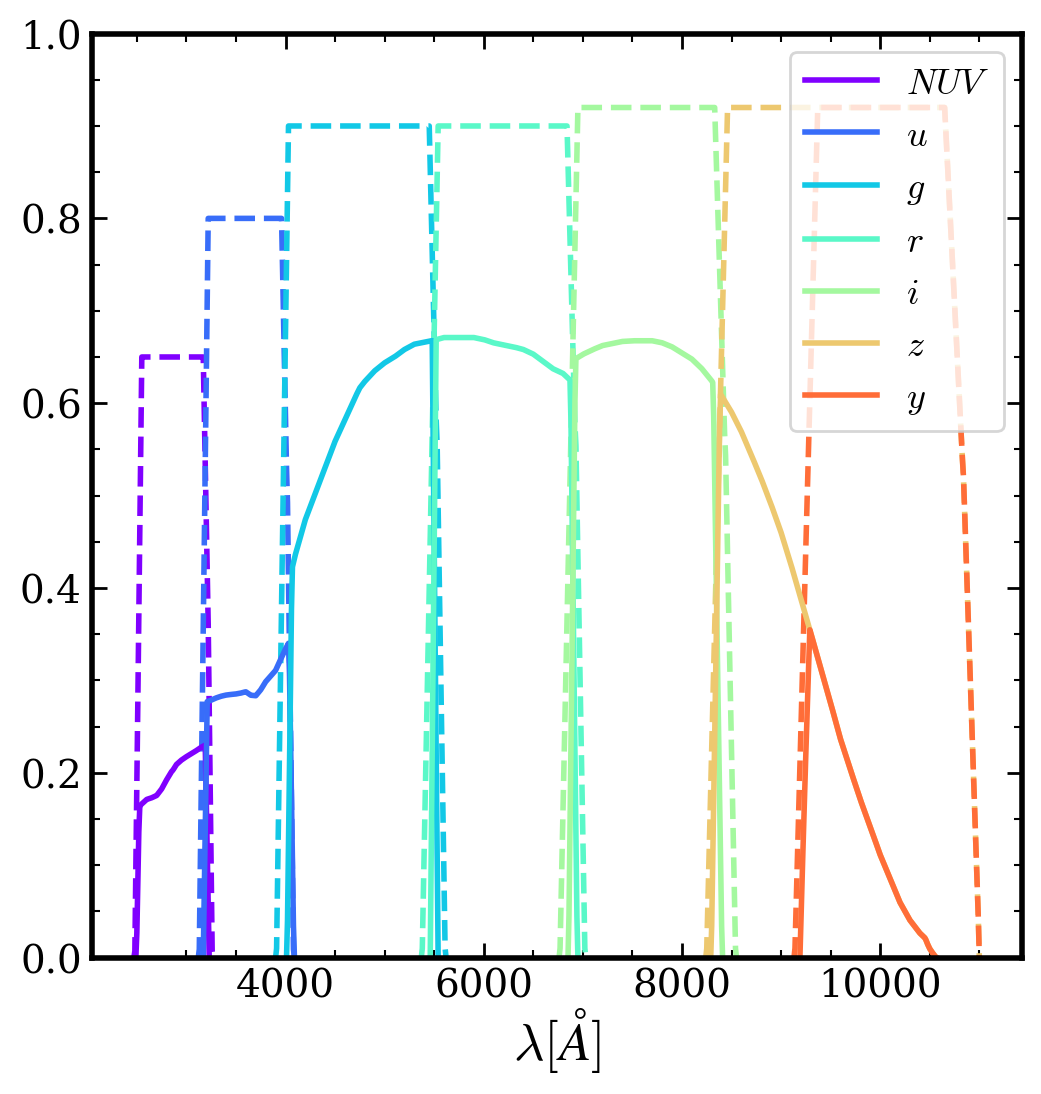}
	\caption{The intrinsic (dashed curves) and real (solid curves) transmissions of the CSST seven photometric filters. The real transmission curves take into account the detector quantum efficiency. The details on parameters of transmissions can be found in \citet{cao2018testing}.}
	\label{fig:filters}
\end{figure}

The Chinese Space Station Telescope (CSST) is a 2-meter space telescope dedicated to the next generation of cosmological observations, scheduled to be launched within the next two years. After launch, the CSST will share its orbit with the Chinese Manned Space Station \citep{zhan2011consideration,cao2018testing,gong2019cosmology}. One of the main tasks of the CSST is the optical survey (CSS-OS), which aims to cover an area of approximately 17,500 square degrees over a period of about 10 years, spanning the optical and near-infrared bands from around 250 nm to 1000 nm. For point sources in the $g$, $r$, and $i$ bands, the 5-sigma limit is approximately 26 AB magnitudes, while for other bands it ranges from 24.5 to 25.5. Figure \ref{fig:filters} illustrates the intrinsic transmissions and real transmissions including detector quantum efficiency of the seven photometric filters employed by CSST. The details of the transmission parameters can be found in \citet{cao2018testing}. The primary scientific objectives of the CSST include studying the evolution of large-scale structures, exploring the properties of dark matter and dark energy, as well as investigating the formation and evolution of galaxies and other areas \citep{gong2019cosmology,cao2022calibrating,zhan2021wide}.

Following, we will briefly explain the process of generating mock data. More detailed information about the mock generation process can be found in \citet{zhou2021spectroscopic}. This mock dataset has been utilized in various studies aligned with the objectives of the CSST, specifically focusing on exploring machine learning and spectral energy distribution fitting techniques \citep{zhou2021spectroscopic,zhou2022photometricBNN,lu2024estimating,luo2024imputation}.

The mock data is designed to exhibit characteristics similar to observations from the CSST survey, including redshift, magnitude distribution, and types of galaxies. To ensure a high level of realism in simulating galaxy images for the CSST photometric survey, we employ mock image generation techniques based on observations taken in the COSMOS field using the Advanced Camera for Surveys of the Hubble Space Telescope (HST-ACS), while incorporating CSST instrumental effects. Galaxy flux data in the mock dataset are derived from these images using aperture photometry. The COSMOS HST-ACS survey covers an area of approximately 2 square degrees in the F814W band, with spatial resolution similar to that of the CSST and an 80\% energy concentration radius of $R_{80}\sim0.^{\prime\prime}15$ \citep{cao2018testing,gong2019cosmology,koekemoer2007cosmos,massey2010pixel,bohlin2016perfecting}. Moreover, the COSMOS HST-ACS F814W survey exhibits significantly lower background noise compared to the CSST survey, expected to be approximately one-third of that experienced in the CSST survey. This attribute provides a solid foundation for simulating CSST galaxy images. Here, we summarize and emphasize the key information regarding the generation of mock data. 

First, we choose a region of 0.85 × 0.85 deg$^2$ from the HST ACS survey, where approximately 192,000 galaxies are detected. Subsequently, we adjust the pixel size from $0.03^{\prime\prime}$ in the HST survey to $0.075^{\prime\prime}$ in the CSST survey. The identified galaxies are then extracted as square stamp images with the galaxies positioned at the center. The dimensions of the images are 15 times the semi-major axis of the galaxies, which can be sourced from the COSMOS weak lensing source catalog \citep{leauthaud2007weak}, resulting in varying sizes of our galaxy images. Furthermore, we mask all sources in the stamp image with a signal-to-noise ratio (SNR) greater than 3$\sigma$, except for the central galaxy, and replace them with the CSST background noise.

We then move on to adjust the scale of the galaxy images from the HST-ACS F814W survey to the CSST flux level. This is accomplished by utilizing galaxy spectral energy distributions (SEDs) to derive the CSST 7-band images. The galaxy SEDs are created by fitting the fluxes and other photometric data from the COSMOS2015 catalog using the LePhare code \citep{arnouts1999measuring,ilbert2006accurate,laigle2016cosmos2015}. In this fitting process, the photo-$z$ values from the catalog are held constant. The 31 SED templates used for the fitting process are also sourced from $LePhare$ code, and are extended from approximately 900Å to around 90Å through the BC03 method \citep{bruzual2003stellar}. This extension allows for the inclusion of fluxes from high-redshift galaxies across all CSST photometric bands. For more detailed information, refer to \citet{cao2018testing}.

We have chosen approximately 100,000 high-quality galaxies with reliable photo-$z$ measurements for the SED fitting procedure. In addition to accounting for dust extinction, we have also taken into consideration emission lines like Ly$\alpha$, H$\alpha$, H$\beta$, [O\uppercase\expandafter{\romannumeral2}], and [O\uppercase\expandafter{\romannumeral3}]. Following the SED fitting, theoretical flux data can be computed by convolving them with the CSST filter transmission curves, illustrated in Figure \ref{fig:filters}. Concurrently, we compute the fluxes of the F814W images using an aperture size of 2 times the Kron radius \citep{kron1980photometry}. The CSST 7-band images are then generated by appropriately rescaling the fluxes. To align with CSST observations, we adjust the background noise to match the same level. More information on the noise adjustment is available in \citet{zhou2022photometricBNN}. Consequently, we obtain mock CSST galaxy images for the seven CSST photometric bands.

For flux measurement in our mock galaxy data, we utilize aperture photometry. Starting with establishing the Kron radius along the major and minor axes, we define an elliptical aperture size of 1 times $R_{\rm{Kron}}$. Within this aperture, we can compute the flux and its corresponding error for each band.

The final CSST mock catalog includes measurements in seven CSST bands, consisting of flux, flux error, and photo-$z$ information for approximately 100,000 galaxies. These CSST mock galaxies were selected from the COSMOS catalog, which utilizes photo-$z$ estimates derived from a comprehensive range of over 30 bands across the electromagnetic spectrum. \citet{laigle2016cosmos2015} conducted a validation process by comparing the photo-z estimates in the COSMOS2015 catalog to various spectroscopic survey samples. Detailed information on the accuracy of the photo-z estimates and the characteristics of spectroscopic redshift (spec-z) samples can be found in Tables 4 and 5, as well as Figures 11 and 12 in the study by \citet{laigle2016cosmos2015}. Given the demonstrated precision and accuracy of the photo-$z$ estimates in the COSMOS catalog, we consider them reliable and have utilized them as the true redshift values, denoted as $z_{\rm true}$ henceforth.

\begin{figure}
	\includegraphics[width=\columnwidth]{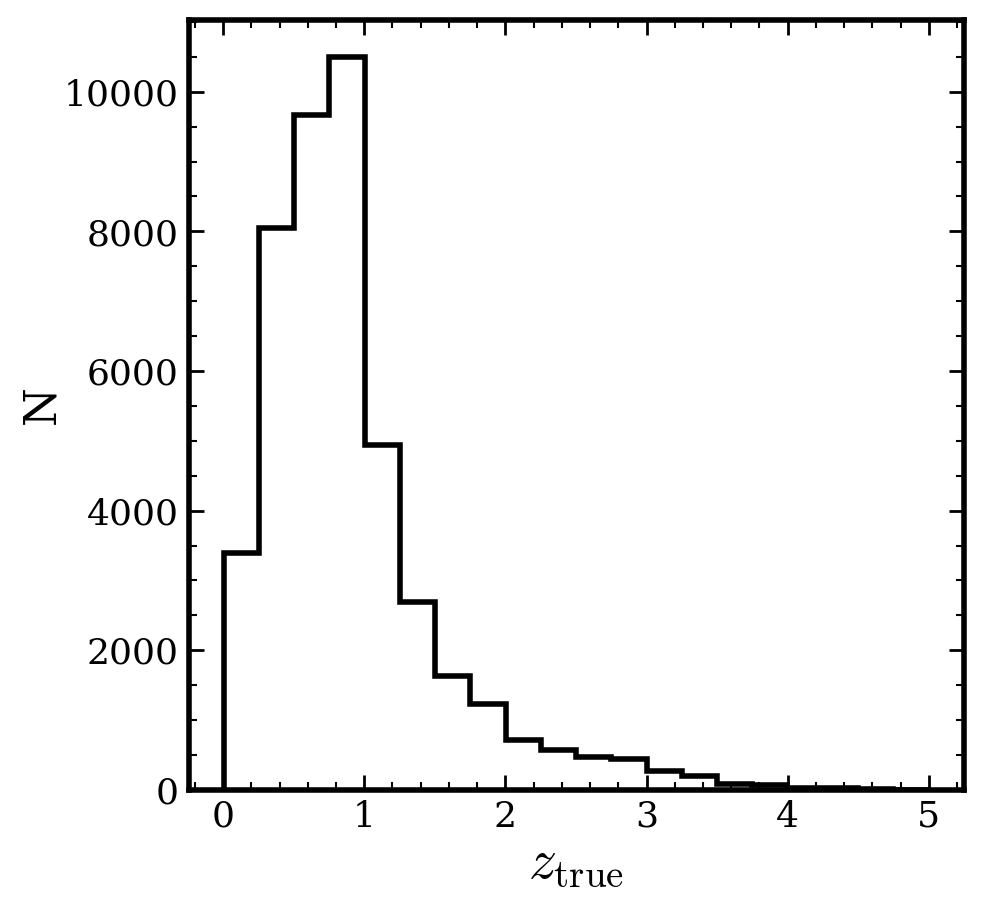}
	\caption{The galaxy redshift distribution of the CSST mock catalog with SNR greater than 10 in g or i bands. The distribution peaks around $z = 0.8 \sim 1.0$, and can reach maximum at $z \sim 5$.}
	\label{fig:redshift}
\end{figure}

From this simulated CSST catalog, we then proceeded to extract a high-quality photometric sub-sample, which comprises sources with a signal-to-noise ratio (SNR) exceeding 10 in either the $g$ or $i$ band. The resulting subsample consists of a total of 44,991 sources, which serves as the primary sample used in our subsequent research. Figure \ref{fig:redshift} illustrates the distribution of redshifts for this dataset, with the majority of sources centered around the range of $z=0.8-1.0$. The redshift distribution extends from 0 to 5, indicating coverage across a wide range of cosmic distances.

\section{Methodology}\label{section:method}

In multi-band observations, different filters correspond to different wavelength coverage ranges, therefore, photometric data can be seen as sequential data arranged in wavelength order. Considering that traditional machine learning models typically struggle with handling sequential data, this study intends to explore a new neural network model,  specifically a LSTM deep learning model, for estimating photo-$z$s. In recent years, LSTM has been increasingly used in astronomical research, but mostly focused on spectral or temporal data analysis  \citep{zhang2018time,hu2022spectroscopic,iess2023lstm,sajad2023modeling,sourav2023predicting}. This study represents the first application of this architecture in the photo-$z$ estimation.

\subsection{Long Short-Term Memory Network (LSTM)}
\label{sec:lstm} 

LSTM, a special type of RNN architecture, was first proposed by \citet{hochreiter1997long} and has since become one of the mainstream models for handling sequence data in subsequent research. Before delving into a detailed description of the LSTM model, let us first provide a brief introduction to RNN.

RNN is a specific type of recursive neural network that takes sequence data as input and recursively evolves along the sequence direction, with all nodes (recurrent units) connected in a chain-like manner. Due to its memory, parameter sharing, and Turing completeness properties, RNN has significant advantages in learning the nonlinear features of sequences. RNN finds wide applications in the field of Natural Language Processing (NLP) such as speech recognition, language modeling, machine translation, and is extensively used in various time series prediction tasks.

Traditional RNNs suffer from the vanishing gradient problem \citep{hochreiter1998vanishing}, especially when dealing with long sequence data. As the number of time steps increases, the gradient of RNN gradually diminishes. This limitation hampers the performance of traditional RNNs in capturing long-term memory and complex dependencies. To address this issue, the LSTM structure was proposed \citep{hochreiter1997long,gers2000learning,graves2013speech,jozefowicz2015empirical}.

Similar to RNNs, LSTM also consists of chain-like modules, but its repeating module structure is more intricate. Each repeat module in LSTM includes a memory cell specifically designed to store long-term information. The memory cell consists of a forget gate, an input gate, an output gate, and a cell state. These four components work together during the time-step iteration process in LSTM to calculate the activation of network units, thereby computing the mapping from input to output sequences.

Specifically, at each time step $t$, the values of the forget gate $f_t$, input gate $i_t$, output gate $o_t$, and the candidate memory cell state $\hat{C}_t$ are computed based on the input sequence data $x_t$ and the hidden state $h_{t-1}$ from the previous time step:
\begin{equation}
     f_t=\sigma\left(W_f \cdot\left[h_{t-1}, x_t\right]+b_f\right),
	\label{eq:forget_door}
\end{equation}
\begin{equation}
     i_t=\sigma\left(W_i \cdot\left[h_{t-1}, x_t\right]+b_i\right),
	\label{eq:input_door}
\end{equation}
\begin{equation}
     o_t=\sigma\left(W_o \cdot\left[h_{t-1}, x_t\right]+b_o\right),
	\label{eq:output_door}
\end{equation}
\begin{equation}
     \tilde{C}_t=\tanh \left(W_C \cdot\left[h_{t-1}, x_t\right]+b_C\right),
	\label{eq:cand_cell_state}
\end{equation}
where the subscript $t$ represents the index of the sequence data (i.e., the time step). $\sigma$ denotes the sigmoid activation function, and $\tanh$ represents the hyperbolic tangent activation function. $W_i$, $W_f$, $W_o$, and $W_C$ correspond to the weight matrices of the input gate, forget gate, output gate, and candidate cell state, respectively. $b_i$, $b_f$, $b_o$, and $b_C$ indicate the bias vectors of the input gate, forget gate, output gate, and candidate cell state, respectively. All weight matrices and bias vectors are reused at each time step.

Subsequently, utilize the values of these gates and the candidate cell state to update the cell state $C_t$ and the hidden state $h_t$:
\begin{equation}
     C_t=f_t \odot C_{t-1}+i_t \odot \tilde{C}_t,
	\label{eq:cell_state}
\end{equation}
\begin{equation}
    h_t=o_t \odot \tanh \left(C_t\right),
	\label{eq:hidden_state}
\end{equation}
where the symbol $\odot$ represents the element-wise multiplication of two vectors. The initial values of the cell state and the hidden state are set to $C_0 = 0$ and $h_0 = 0$, respectively.

Throughout the entire process, the forget gate determines which information to discard from the memory cell of the previous time step, the input gate controls the input of new information, and the output gate regulates the flow of output information. This gate-controlled mechanism enables LSTM networks to effectively handle sequence data while maintaining long-term memory and efficiently capturing crucial dependencies within the sequence, thereby improving model performance and accuracy. Through the structure and gate control mechanism of LSTM, the vanishing gradient problem faced by traditional RNNs when dealing with long sequence data has been successfully addressed, making it a robust tool for handling long-term dependencies in sequence data.

\subsection{Model Architectures for Photo-$z$}
\label{sec:model_archi} 

\begin{figure}
    \centering
	\includegraphics[width=0.78\columnwidth]{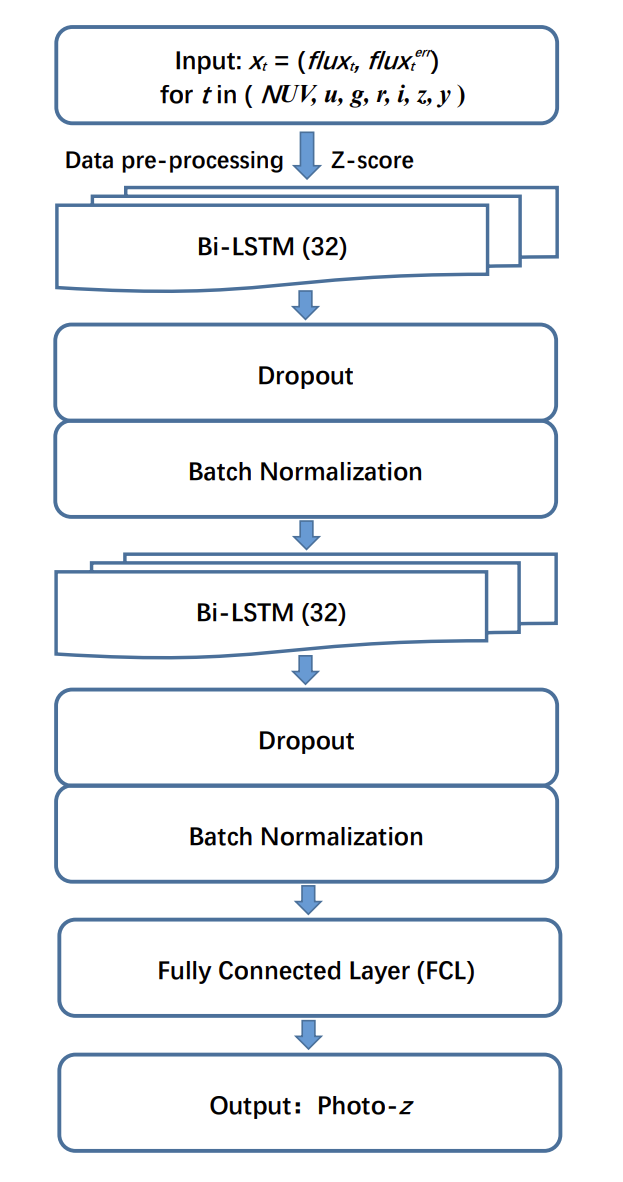}
	\caption{The architecture of the proposed LSTM model. The top boxes represent the input layer, where the input data is a sequence of photometric data from various bands of CSST. The two boxes labeled Bi-LSTM represent bidirectional LSTM modules, each containing 32 neurons and connected to a Dropout layer and a Batch Normalization layer. The final Batch Normalization layer is connected to a Fully Connected Layer (FCL) for photo-$z$ output.}
	\label{fig:frame_lstm}
\end{figure}

The aim of this study is to estimate the photo-$z$s of galaxies observed by CSST. This can be considered as a photo-$z$ regression prediction problem derived from the sequence of photometric data, as the photo-$z$ of galaxies depends on the photometry in various bands.The network architecture designed to address this regression problem is depicted in Figure \ref{fig:frame_lstm} and implemented using the Python libraries Keras \footnote{\url{https://keras.io}} and TensorFlow 2 \citep{abadi2016tensorflow}. 

We first combine the galaxy photometric flux $flux_t$ and error $flux^{err}_t$ in each band into an array ($flux_t, flux^{err}_t$), and arrange them into a sequence from short to long wavelengths to serve as input for the model. Before detailing the specifics of our LSTM model, we first performed some pre-processing on the input data. This is important because if the variation in the input data is large, it may negatively impact the model’s learning ability \citep{Jain2000Statistical,2006Data}. To ensure that all input variables are on the same scale, we standardized the input sequence to ensure the convergence stability of the model parameters developed in this study. The standardization formula is as follows:
\begin{equation}
    x^*=\frac{x-\mu}{\sigma},
	\label{eq:input_standardized}
\end{equation}
where $x$ represents the value of each input variable, while $\mu$ denotes the mean and $\sigma$ indicates the standard deviation of all data across these variables.

The proposed deep learning network utilized two bidirectional LSTM layers (Bi-LSTM) \citep{schuster1997bidirectional}, each with 32 hidden units, followed by a dropout layer and Batch Normalization layer. Bi-LSTM integrates information from both the forward (from the start to the end of the sequence) and backward (from the end to the start of the sequence) directions to enhance the learning capabilities of model and improve the robustness of neural network. After inputting the data, the LSTM can reorganize the input data and retain previous information, aiding in improving the ability of model to learn from wavelength sequence data. The dropout layers added after each Bi-LSTM layer help prevent overfitting. Dropout is an efficient regularization technique that randomly drops out some elements of the weight matrix with a certain probability, suppressing overfitting between neurons \citep{hinton2012improving,zaremba2014recurrent,srivastava2014dropout}. This helps reduce the reliance of model on training data, encourages more balanced learning between neurons, and improves generalization ability. During the model prediction phase, dropout can also be enabled (Monte Carlo dropout), giving the network a level of stochasticity \citep{gal2016dropout,gal2016theoretically}. Through Monte Carlo dropout, even after training, the model is no longer deterministic. This is achieved by using dropout masks as the implementation of a Bernoulli process. By estimating the variance of the prediction results obtained through multiple forward passes, we can assess the uncertainty of model and obtain a probability density distribution (PDF) of predicted photo-$z$s. In our study, we set the dropout rate for both bidirectional LSTM layers to 0.15. Additionally, the Batch Normalization layers added after each Dropout layer can help reduce overfitting problems \citep{ioffe2015batch}, normalize the input of each mini-batch, accelerate the training process, and assist the network in faster convergence.

Finally, we included a fully connected layer (FCL) after the second Batch Normalization layer to enhance the model learning capabilities and generate predictions for photo-$z$s. Given that our problem is a regression task, we employed the Mean Squared Error (MSE) loss function as follows:
\begin{equation}
    L O S S=\sum_{i=1}^{\mathrm{N}}\left(\mathrm{y}_{i}-\hat{\mathrm{y}}_{i}\right)^2,
	\label{eq:loss_func}
\end{equation}
where $\mathrm{y}_{i}$ is the observed value of the $i$-th sample, and $\hat{\mathrm{y}}_{i}$ is the predicted value of the $i$-th sample. We utilized the Adam optimizer \citep{bae2019does,mehta2019cnn} to optimize the network weights. This optimizer adaptively adjusts the learning rate for each weight parameter based on the gradients and squared gradients calculated in the past. For our model, we set the initial learning rate to 0.0001.

\section{ Model Applied to CSST Photo-z Estimations}

In this section, we will apply the proposed LSTM model to the simulated data from the Chinese Space Station Telescope (CSST) to validate its capability in estimating photometric redshifts. Given that the photometric redshift data from the future CSST will be primarily used for weak gravitational lensing surveys, the demand for accuracy is extremely high. We aim for our research findings to advance the anticipated scientific investigations associated with the future CSST.


\subsection{Model Evaluation Criteria}

Here, we used three metrics to assess the quality of the model's photo-$z$ estimation: the catastrophic outlier rate ($f_{out}$), normalized median absolute deviation ($\sigma_{\mathrm{NMAD}}$) and photometric redshift bias ($bias$).

The catastrophic outlier rate ($f_{out}$) is used to identify exceptional cases in photo-$z$ estimation. Sources whose photo-$z$ estimates satisfy the following condition \citep{fotopoulou2018cpz,hildebrandt2010phat,euclid2023selection},
\begin{equation}
	\frac{|z_{\rm true}-z_{\rm phot}|}{1+z_{\rm true}}>0.15,
\end{equation}
are considered catastrophic outliers, meaning their photo-$z$s are incorrect, where $z_{\rm true}$ is the reference redshift used as the "ground truth" and $z_{\rm phot}$ is the predicted photo-$z$. This metric is widely applied in the evaluation of photo-$z$s. For example, the PHoto-z Accuracy Testing programme (PHAT) utilizes this outlier rate to assess the accuracy of various photo-$z$ methods \citep{hildebrandt2010phat}.

The $\sigma_{\mathrm{NMAD}}$ is utilized to measure the accuracy of photo-$z$ estimation and is defined as follows \citep{brammer2008eazy}:
\begin{equation}
	{\rm \sigma_{NMAD}}=1.48\times {\rm median}\left(\left|\frac{\Delta z - {\rm median}(\Delta z)}{1+z_{\rm true}}\right|\right),
\end{equation}
 where 
\begin{equation}
	\Delta z =z_{\rm phot}-z_{\rm true}.
\end{equation}
We opted for $\sigma_{\mathrm{NMAD}}$ over standard deviation due to its lower sensitivity to outliers and incorporation of a scaling factor of 1.48, allowing $\sigma_{\mathrm{NMAD}}$ to be interpreted as the standard deviation of normally distributed data.

The $bias$ refers to the deviation of photometric redshift and is used to quantify the systematic discrepancy between our measured values of galaxy redshifts and their true values. Typically, the following formula is used to calculate this:
 \begin{equation}
	bias = {\rm median}\left(\frac{z_{\rm true}-z_{\rm phot}}{1+z_{\rm true}}\right).
\end{equation}
This metric allows us to assess whether and to what extent we are systematically overestimating or underestimating the redshifts of galaxies.

\subsection{Photo-$z$ Results}

In Section \ref{section:mock}, we selected approximately 45,000 high-quality sources with signal-to-noise ratios greater than 10 in the $g$ or $i$ bands. These sources all have flux and flux error data, and high-quality photo-$z$ measurements in the COSMOS catalog, which can be considered as true redshifts. Our model will be trained and tested within this sample. For future CSST surveys, we will use true spect-$z$ samples obtained from future spectroscopic surveys as training samples.

\begin{figure}
    \centering
	\includegraphics[width=\columnwidth]{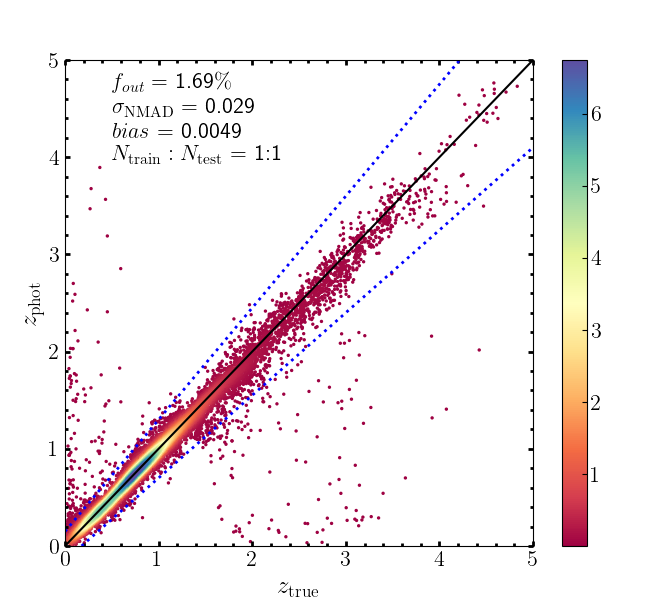}
	\caption{True redshifts versus predicted photo-$z$ derived from the proposed LSTM model. $f_{out}$, $\sigma_{\mathrm{NMAD}}$, and $bias$ denote the catastrophic outlier rate, normalized median absolute deviation, and photometric redshift bias, respectively. $N_{\rm{test}}$ and $N_{\rm{train}}$ represent the number of testing and training samples. The solid line illustrates the diagonal, while the dashed line indicates $|z_{\rm{true}} - z_{\rm{phot}}|/(1 + z_{\rm{true}}) = 0.15$. The colors signify the density of points.}
	\label{fig:accuracy_z_1_1}
\end{figure}

The approximately 45,000 galaxy samples were divided into training and testing sets. We randomly selected 22,500 samples for training, while the remaining 22,491 were allocated for testing, resulting in an approximate training-to-testing ratio of 1:1. The results of our LSTM model for photo-$z$ estimation are shown in Figure \ref{fig:accuracy_z_1_1}. From the figure, it is evident that the photo-$z$ of the test samples exhibits a high level of accuracy, with catastrophic outlier rate ($f_{out}$), normalized median absolute deviation ($\sigma_{\mathrm{NMAD}}$), and photometric redshift bias ($bias$) being 1.69\%, 0.029, and 0.0049, respectively, demonstrating the robust capabilities of the proposed LSTM model in CSST photo-$z$ estimation.

On the other hand, when utilizing our model for photo-$z$ prediction within the CSST context, the activation of MC dropout allows for the generation of statistical results for redshift prediction. By conducting a detailed analysis of these results, a redshift probability density function (PDF) can be developed, highlighting a significant capability of our proposed LSTM model.

\citet{carrasco2014som} and \citet{rau2015accurate} have investigated various methods for generating probability density functions (PDFs), such as Gaussian Mixture Models (GMM) and non-parametric techniques like Kernel Density Estimation (KDE) \citep{murray1956remarks,parzen1962estimation}. In this study, we opted to utilize KDE for constructing our PDF. The KDE method involves placing a kernel at each data point, typically a smooth bell-shaped function, and aggregating these kernels to form a continuous smooth probability density approximation; detailed computational methodologies can be found in \citet{lu2024estimating}. Throughout the computation process, we employed the $scipy.stats.gaussian\_kde$\footnote{\url{https://docs.scipy.org/doc/scipy/reference/generated/scipy.stats.gaussian_kde.html}} class with all parameters set to default configurations. Additionally, $sklearn.neighbors.KernelDensity$\footnote{\url{https://scikit-learn.org/dev/modules/generated/sklearn.neighbors.KernelDensity.html}} also provides a flexible way to perform kernel density estimation and supports Gaussian kernel functions. It is essential to clarify that our PDF does not represent a conventional probability density function but rather signifies a predictive distribution. 

\begin{figure*}
    \centering
	\includegraphics[width=\textwidth]{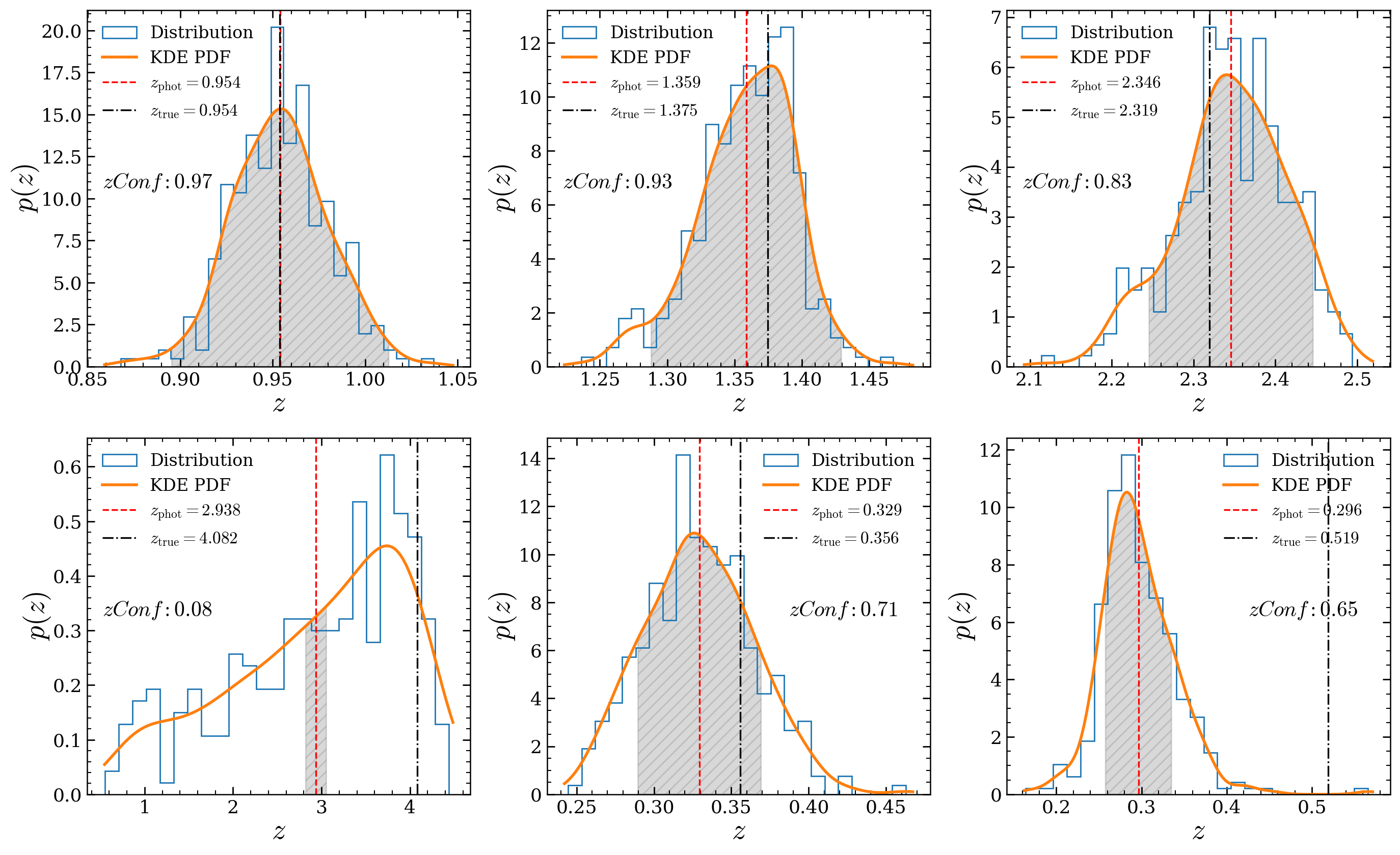}
	\caption{Examples of the probability density distribution (PDF) of predicted redshift by the proposed LSTM model. The histogram represents the statistical distribution of predicted redshifts. The solid line represents the KDE PDF. The vertical dash-doted line corresponds to the true redshift $z_\mathrm{true}$. The vertical dashed line represents the model-predicted photometric redshift $z_\mathrm{phot}$.The grey area under the KDE PDF is $zConf$.}
	\label{fig:zpdf}
\end{figure*}

We randomly selected six galaxies to showcase their probability density functions (PDFs), as shown in Figure \ref{fig:zpdf}. Since the true distributions are not available, we cannot directly validate the PDFs of these galaxies. To assess the overall effectiveness of the PDFs generated by the LSTM model, it is necessary to conduct statistical studies on the PDFs of the galaxies in the test samples.

The performance of the model-generated PDFs can be evaluated using various metrics. This paper adopts a widely used standard method from previous research, probability integral transform (PIT) plots, to assess the agreement between the predicted distributions and the true uncertainties \citep{dawid1984present,d2018photometric,desprez2020euclid,mucesh2021machine}. For each galaxy, the PIT is calculated as the cumulative distribution function (CDF) at the true redshift:
\begin{equation}
PIT=\int_{-\infty}^{z_{\rm{true }}} PDF(z) dz, 
\end{equation}
where $z_{\rm{true}}$ represents the true redshift, and $PDF(z)$ is the probability density function PDF at redshift $z$.

\citet{mucesh2021machine} point out that for the entire sample or any sufficiently large galaxy sample, a well-calibrated PDF should generate PIT values that follow a uniform distribution, specifically a standard uniform distribution ($U(0, 1)$; \citealt{dawid1984present}). This characteristic arises because if the PDF is probabilistically calibrated, the true redshift should be regarded as a random sampling from its distribution. Consequently, the CDF related to the true redshift should not exhibit any preferential values.

If the shape of the PIT distribution deviates from uniformity, this may indicate systematic bias in the inferred PDF distribution. When the PIT distribution appears concave or convex, the PDF typically shows under-dispersion or over-dispersion, respectively. Moreover, changes in the slope of the PIT distribution suggest that there may be systematic errors in the estimates \citep{polsterer2016uncertain}. The PIT distribution has been widely used to validate the effectiveness of redshift PDFs \citep{bordoloi2010photo,polsterer2016uncertain,10.1093/pasj/psx077,desprez2020euclid,mucesh2021machine}.

\begin{figure}
    \centering
	\includegraphics[width=\columnwidth]{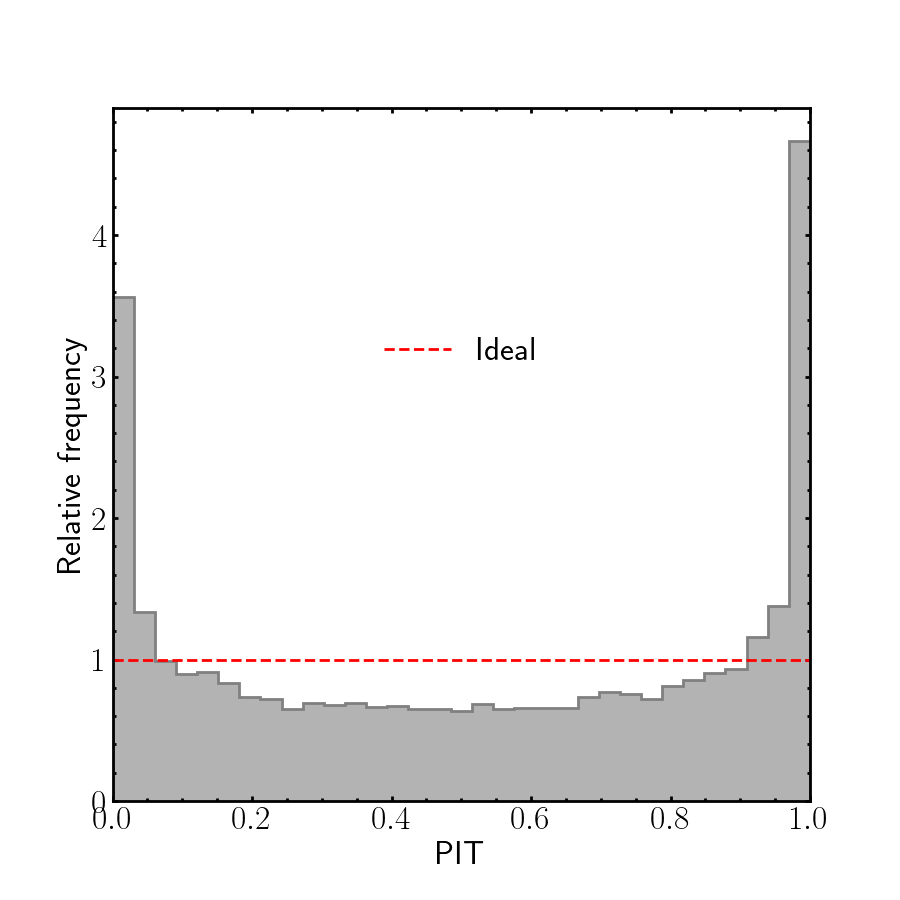}
	\caption{Probability integral transform (PIT) distribution for the proposed LSTM model.The gray solid bars represent the relative frequencies of the test dataset calculated from the PDF predictions. The red horizontal dashed line indicates the ideal uniform distribution, representing the expected distribution characteristics that a well-calibrated probability density function should follow.}
	\label{fig:pit}
\end{figure}

Figure \ref{fig:pit} shows the PIT plot generated by our proposed LSTM model. Overall, the PIT distribution of the LSTM model appears to exhibit uniform characteristics, indicating that the estimates of photo-$z$ do not exhibit significant systematic overestimation or underestimation. However, peaks at the edges suggest that the PDFs produced by the LSTM model have difficulty adequately explaining certain outliers and may also suffer from under-dispersion, meaning the generated PDFs are too narrow. Compared to most methods tested in the Euclid photometric redshift challenge (see Figure 7 in \citealt{desprez2020euclid}), the performance of the LSTM model is quite satisfactory. However, it performs relatively poorly compared to the methods tested by \citet{mucesh2021machine} on the Dark Energy Survey (DES) sample. Since our model-generated PDFs are obtained through the use of Monte Carlo (MC) dropout techniques during the prediction phase, they focus more on the model’s predictive distribution rather than the true observed distribution in a statistical sense. Treating the photo-$z$ estimation as a classification problem or employing a Mixture Density Network (MDN; \citealt{bishop1994mixture}) at the end of the fully connected layer of the prediction model may provide a more effective approach for deriving PDFs \citep{teixeira2024photometric}.

After obtaining the PDF, we can introduced a new metric called $zConf$ to quantify the confidence level of the predicted redshift \citep{carrasco2013tpz}. In this study, we define $zConf$ as the integral of the PDF within the range $z_\mathrm{phot} \pm \alpha(1 + z_\mathrm{phot})$, where $z_\mathrm{phot}$ represents the mean value of the PDF. We choose $\alpha = 0.03$ to approximate the intrinsic scatter of the algorithm when implemented on the data.
Figure \ref{fig:zpdf} displays the $zConf$ values for each PDF. In these examples, galaxies with unimodal or concentrated PDFs typically exhibit higher $zConf$ values, with the peak more likely to be near the true redshift. Conversely, galaxies with bimodal or dispersed PDFs tend to have lower $zConf$ values, indicating that $zConf$ can provide a reasonable confidence level for the redshift estimation. It is worth noting that a lower $zConf$ does not necessarily imply that the prediction itself is inaccurate or of low quality. Instead, it reflects the level of uncertainty associated with the prediction, which is based on the underlying model and input data. In some cases, the model may generate low-confidence predictions due to limited data, high variability in the observed data, or the complexity of the underlying relationships. However, even if the model is uncertain about these predictions, the actual outcomes may still be reasonable and valuable. The $zConf$ metric reflects the shape of the PDF rather than its accuracy.

\begin{figure}
    \centering
	\includegraphics[width=\columnwidth]{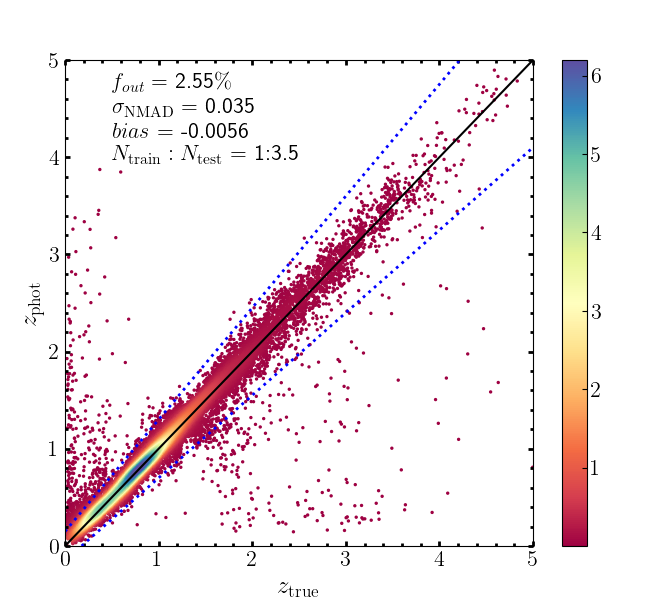}
	\caption{Same as Fig. \ref{fig:accuracy_z_1_1}, but corresponding to a 1:3.5 training to testing ratio.}
	\label{fig:accuracy_z_1_35}
\end{figure}

\begin{figure}
    \centering
	\includegraphics[width=\columnwidth]{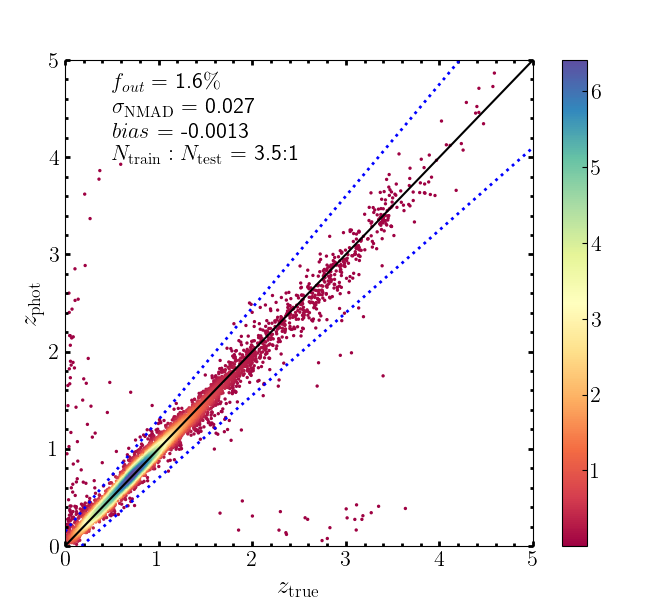}
	\caption{Same as Fig. \ref{fig:accuracy_z_1_1}, but corresponding to a 3.5:1 training to testing ratio.}
	\label{fig:accuracy_z_35_1}
\end{figure}

To investigate the impact of the quantity of training data on photo-$z$ estimation, we also experimented with training-to-testing ratios approximately 1:3.5 and 3.5:1, with the results presented in Figure \ref{fig:accuracy_z_1_35} and \ref{fig:accuracy_z_35_1} respectively. From the figures, it is evident that with the increase in training data, the accuracy of photo-$z$ estimation also gradually improves. For the scenario with a training-to-testing ratio of 3.5:1, the catastrophic outlier rate ($f_{out}$), normalized median absolute deviation ($\sigma_{\mathrm{NMAD}}$), and photometric redshift bias ($bias$) in the test set achieved values of 1.6\%, 0.027, and -0.0013; respectively.

\subsection{Comparison with Other Methods}

We compared the results of the proposed LSTM model with three other methods: the Easy and Accurate Zphot from Yale software (EAZY) \citep{brammer2008eazy}, the Weighted Random Forest algorithm (WRF) \citep{lu2024estimating}, and the Multi-layer Perceptron Photometric Redshift Code (MLP) \citep{zhou2021spectroscopic}. To better understand the effectiveness and applicability of these methods, as well as the configurations they utilize, we will first summarize their basic features here, including their working principles, input parameters, and algorithm characteristics.

EAZY is one of the most widely used software among all template-fitting methods for photo-$z$ estimation. As described by \citet{brammer2008eazy}, it estimates galaxy redshifts using medium and broad-band photometry data, involving the optimization of the differences between observed values and fluxes based on SED templates through minimizing the chi-square statistic. Several studies have shown that EAZY exhibits outstanding performance in photo-$z$ estimation \citep{yang2014photometric,chen2018xmm}, and \citet{desprez2020euclid} demonstrated that template-fitting methods provide nearly identical results when run in identical configurations. The differences observed in the results are not due to differences in performance of the template-fitting methods, but rather variations in their configurations. Therefore, we chose the EAZY method as a representative template-fitting approach for comparison with our LSTM algorithm.  In the EAZY results, we utilized the standard CWW+KIN template set, which is based on the empirical CWW template set \citep{2003Colors} with extensions prescribed by \citet{Kinney1996Template}. This template set, comprised of six templates, is widely used in photo-$z$ estimation. Additionally, we incorporated the $r$ band apparent magnitude prior $p(z|m_r)$, representing the redshift distribution of galaxies with apparent magnitude $m_r$. The Z\_STEP\_TYPE parameter was set to 0, indicating a evenly spaced redshift grid. All other parameters were retained at their default settings. The input bands consisted of the seven CSST bands described before.

The WRF method utilizes Random Forest (RF) as a machine learning algorithm to estimate photometric redshifts. This method uses galaxy flux and color as input features and establishes a mapping between flux and redshift by employing a training dataset that includes available spectroscopic redshifts. In order to enhance the accuracy and reliability of photo-$z$ predictions, the WRF method trains the model by weighting and perturbing samples based on flux errors. Research results indicate that the WRF method achieves a significantly high level of accuracy in photo-$z$ estimation. \citet{lu2024estimating} utilized WRF to estimate photo-$z$s on the same sample utilized in this study. For comparison purposes, we will redivide the training and testing datasets according to a uniform standard and harmonize the input features to ensure fairness and comparability of the results.

MLP is a commonly used neural network model widely employed for photometric redshift estimation. Comprising multiple layers of neurons, MLP consists of an input layer, hidden layers, and an output layer. By iteratively optimizing network parameters on the training dataset, the MLP model can learn the complex nonlinear mapping relationship between input features and actual redshift values, enabling accurate estimation of photo-$z$s. In this study, we utilized the MLP model architecture proposed by \citet{zhou2021spectroscopic} to compute photo-$z$s for CSST. This architecture includes two hidden layers, each containing 40 neurons, with ReLU activation function applied after each hidden layer. The design of the number of neurons in each layer adheres to the classic MLP structure format: $n: 2n: … 2n: 1$, where $n$ represents the number of input data elements. Our MLP model has a total of 20 inputs, including seven flux values, seven error values, and six color features. Regarding the number of hidden layers, we found that two hidden layers are sufficient to achieve accurate photo-$z$ estimation for CSST. Increasing the number of hidden layers does not significantly improve the prediction accuracy; instead, it may reduce the computational efficiency of the model and increase the risk of overfitting, as confirmed by the research of \citet{zhou2022photometricBNN}. For more specific details, please refer to Figure 8 and Table 2 provided by \citet{zhou2021spectroscopic}.

\begin{table}
	\centering
	\caption{Results of photo-$z$ estimation for CSST mock data using four different methods. $f_{out}$, $\rm \sigma_{NMAD}$, and $bias$ with superscripts "a" and "b" respectively represent the results obtained by models using different input features. Superscript "a" corresponds to input features consisting of fluxes and errors of the 7 bands ($NUV$, $u$, $g$, $r$, $i$, $z$, $y$) of CSST; whereas superscript "b" corresponds to input features that include not only the fluxes and flux errors of the 7 bands of CSST, but also 6 colors ($NUV-u$, $u-g$, $g-r$, $r-i$, $i-z$, $z-y$).}
	\label{tab:zresults}
	\resizebox{\linewidth}{!}{
	\begin{tabular}{ccccc} 
		\hline
            \hline
		 \textbf{Metric} & \textbf{LSTM} & \textbf{WRF} & \textbf{MLP} & \textbf{EAZY}\\
		\hline
          \multicolumn{5}{c}{\textbf{training-to-testing ratio: 1:3.5}}\\
            \hline
		${f_{out}}^a$ &2.55\%  & 3.86\%  & 7.52\%  & 4.13\%\\
		${f_{out}}^b$ & - & 2.34\% & 5.28\% & -\\
            \hline
		${\rm \sigma_{NMAD}}^a$ & 0.035 & 0.039  & 0.053 & 0.041\\
		  ${\rm \sigma_{NMAD}}^b$ & - & 0.027 & 0.044 & -\\
		\hline
		$bias^a$  & 0.0056   & 0.0011 &0.0018 & 0.0064\\
            $bias^b$ & - & 0.0009 & -0.0007 & -\\
            \hline
            \hline
          \multicolumn{5}{c}{\textbf{training-to-testing ratio: 1:1}}\\
            \hline
		${f_{out}}^a$ &1.69\%   & 2.97\%  & 5.02\% &  4.11\%\\
		${f_{out}}^b$ & - & 2.06\% & 3.92\% & -\\
            \hline
		${\rm \sigma_{NMAD}}^a$ & 0.029 & 0.034  & 0.043 & 0.040\\
		  ${\rm \sigma_{NMAD}}^b$ & - & 0.025 & 0.037 & -\\
		\hline
		$bias^a$ & 0.0049   & 0.0014 &0.0008 & 0.0065\\
            $bias^b$ & - & 0.0009 & 0.0002 & -\\
            \hline
            \hline
          \multicolumn{5}{c}{\textbf{training-to-testing ratio: 3.5:1}}\\
            \hline
		${f_{out}}^a$ &1.60\%  & 2.79\%  & 3.67\% &  4.41\% \\
		${f_{out}}^b$ & - & 2.03\% & 3.40\% & -\\
            \hline
		${\rm \sigma_{NMAD}}^a$ & 0.027 & 0.032  & 0.038 & 0.041\\
		  ${\rm \sigma_{NMAD}}^b$ & - & 0.025 & 0.036 & -\\
		\hline
		$bias^a$ & 0.0013   & 0.0006 & -0.0102 & 0.0064\\
            $bias^b$ & - & 0.0005 & -0.0032 & -\\
		\hline
            \hline 
            \end{tabular}
            }
\end{table}

Table \ref{tab:zresults} presents the results of photo-$z$ estimation for simulated CSST data using the four different methods. The table clearly shows that the LSTM model achieves the highest accuracy in photo-$z$ estimation when utilizing only photometric flux and error as input features. Across various ratios of training and testing data, the LSTM model displays a notable advantage in two out of three evaluation metrics for assessing photo-$z$ quality: $f_{\mathrm{out}}$ and $\sigma_{\mathrm{NMAD}}$. For example, in the scenario with a training-to-testing ratio of 1:1, the LSTM model shows a reduction of approximately 60\% and 40\% in $f_{\mathrm{out}}$ compared to the EAZY method and WRF model, respectively, and is only one-third of that of the MLP model. Furthermore, $\sigma_{\mathrm{NMAD}}$ is reduced by approximately 15-30\% compared to the other three methods.

Upon further analysis of Table \ref{tab:zresults}, incorporating not just photometric flux and error but also colors as input features, totaling 20 features (7 bands of fluxes, 7 bands of flux errors, and 6 colors - $NUV-u$, $u-g$, $g-r$, $r-i$, $i-z$, $z-y$), results in a significant improvement in photo-$z$ estimation accuracy for the CSST dataset, especially for the WRF method. This highlights the importance of feature selection in these machine learning models, indicating that skillful combination of input features through proper feature engineering can enhance the quality of photo-$z$ estimation. Nevertheless, despite the inclusion of colors as inputs, in most cases, both the WRF and MLP methods still show relatively lower accuracy in photo-$z$ estimation compared to the proposed LSTM model as seen in Table \ref{tab:zresults}.

Additionally, the results in Table \ref{tab:zresults} demonstrate that across all methods, the deviations $bias$ obtained from photo-$z$ estimation for all samples are minimal, indicating minimal systematic biases when estimating redshifts for galaxies.

\begin{figure}
    \centering
	\includegraphics[width=\columnwidth]{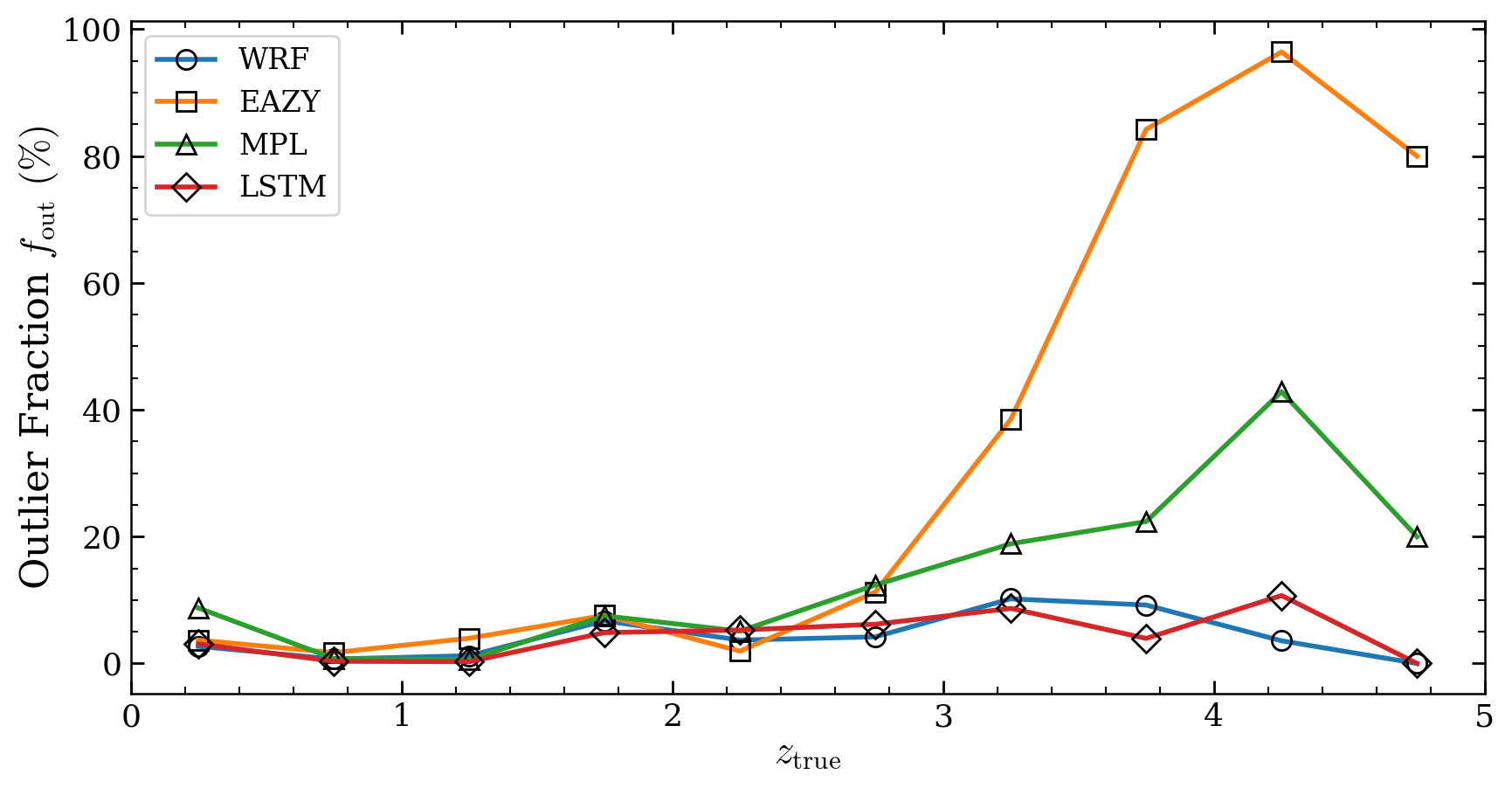}
	\caption{Outlier fractions in different $z_\mathrm{true}$ bins. The triangle, square, circle, and diamond respectively indicate the mean values of the whole data in the current bins for the MLP, EAZY, WRF, and the proposed LSTM models.}
	\label{fig:redshift_bin_results}
 \end{figure}

Furthermore, a detailed comparison was conducted among these four methods. Initially, samples were grouped based on redshift and $f_{out}$ was calculated for each group. The results of this grouping calculation are depicted in Figure \ref{fig:redshift_bin_results}, with a training-to-testing ratio of 1:1 and the inclusion of colors as inputs for both the MLP and WRF models. 

Starting with the comparison between the MLP and LSTM models, Figure \ref{fig:redshift_bin_results} illustrates that within the redshift range of approximately 0.7 to 2.3, the difference in $f_{out}$ between the MLP method and the LSTM model is minimal. This is because, within this range, the colors of adjacent bands in the CSST data effectively capture crucial features of the galaxy SED shape, such as the Balmer/4000Å break or Lyman break features, enabling the color-included MLP method to achieve high accuracy in photo-$z$ estimation. However, beyond this redshift range, especially at higher redshifts, the colors of adjacent CSST bands are not sufficient to fully represent the shape of the galaxy SED, with the Balmer/4000Å break feature moving out of the observable range. This results in the MLP model significantly lagging behind the LSTM model, even with color included. The LSTM model, with its unique memory cell structure, can effectively capture the dependencies among the flux data in different bands, leading to a better understanding of the galaxy’s SED shape.

Moreover, Figure \ref{fig:redshift_bin_results} indicates a notable increase in $f_{out}$ using the EAZY method after the redshift surpasses 2.3. This could be attributed to the fitting templates employed by the EAZY method, which are based on observations of nearby galaxies and may not perform optimally in high-redshift scenarios. Furthermore, the WRF method demonstrates nearly identical $f_{out}$ values compared to the LSTM method across different redshift intervals, emphasizing the robust performance of this method in photometric redshift estimation.

In conclusion, organizing multi-band photometric data into wavelength-ordered sequences for input allows our LSTM network model to autonomously learn and capture hidden patterns and correlations within the data sequences. By effectively handling the complex relationships between different bands, the model achieves accurate predictions of photometric redshifts.

\section{Summary and Discussion}

In this study, we have introduced a novel approach for estimating the photometric redshifts (photo-$z$) of galaxies. Our method treats the photometric fluxes of galaxies in each band as sequential data ordered by wavelength. Recognizing the limitations of traditional machine learning models in handling sequential data, we have chosen to utilize a deep learning model based on a recurrent neural network (RNN), specifically the Long Short-Term Memory (LSTM) architecture, for photo-$z$ estimation, marking the first application of this architecture in this particular domain.

Previous studies have employed various machine learning models, such as Random Forests (RFs) and Multilayer Perceptrons (MLP), for photo-$z$ estimation. While these models performed well with disordered data, they encountered difficulties in processing sequential data, struggling to effectively capture patterns and correlations within the sequence. Therefore, careful feature selection is crucial when utilizing these models for photo-$z$ prediction. Typically, this involves using photometric fluxes from different bands and color information between neighboring bands as inputs to the model. More sophisticated techniques involve developing new machine learning models, such as utilizing decision tree frameworks and ensemble learning methods like Adaboost, to choose the most appropriate features from all available photometric and derived data for accurate photo-$z$ prediction.

In recent years, Convolutional Neural Networks (CNNs) have become widely used in photo-$z$ estimation. While CNNs do not require feature selection, their design is primarily suited for processing image data, requiring the input of galaxy images in different bands for learning within the network. This necessity demands substantial computational resources and storage space, making CNNs more suitable for predicting on smaller-scale datasets.

Our proposed LSTM model effectively addresses various challenges faced by previous models, automatically learning the correlations within the photometric data without complex feature selection requirements. Additionally, our model does not necessitate extensive processing of image data, reducing computational demands and improving efficiency in photo-$z$ estimation tasks, rendering it suitable for situations where only photometric flux data from different bands is available.

The LSTM model we have developed incorporates two LSTM layers, enhancing its ability to understand and model photometric sequence data. This allows for the effective capture of intricate relationships within the data and better identification of patterns and trends. By inputting photometric fluxes from various bands of celestial objects, our model achieves accurate photo-$z$ estimation. Furthermore, by employing the Monte Carlo dropout technique during prediction, our model can generate Probability Density Functions (PDFs) for the predicted redshifts. To evaluate the confidence of predicted redshifts, we introduce a metric named $zConf$, which holds significant importance in cosmological studies.

We applied our method to simulated data from the Chinese Space Station Telescope (CSST) and compared it with three other methods: EAZY, WRF, and MLP. The simulated CSST data was derived from HST-ACS and the COSMOS catalog, taking into consideration future CSST instrument effects. Our LSTM model demonstrated substantial advantages in predicting photo-$z$ compared to the other methods, exhibiting remarkable enhancements in outlier fraction $f_{out}$ and normalized median absolute deviation $\rm \sigma_{NMAD}$. This study presents a new approach for accurately estimating the photo-$z$ of galaxies using photometric data from large-scale survey projects.

We note that after the submission of this paper, \citet{teixeira2024photometric} employed a different RNN architecture, namely the Legendre Memory Unit (LMU; \citealt{voelker2019legendre}), in conjunction with a Mixture Density Network (MDN; \citealt{bishop1994mixture}) to estimate photometric redshifts and derive the probability density functions (PDFs). This study utilized real data from the DECam Local Volume Exploration Survey (DELVE; \citealt{drlica2022decam}) and obtained accurate estimates of photo-$z$s along with their corresponding PDFs. This indicates that the application of RNNs in photo-z estimation is starting to attract attention.

It is important to note that our study utilized photometric fluxes and errors from galaxies in various bands as the input sequence for the model. In future research, the incorporation of additional photometric parameters, such as morphological features from each band, could expand the dimensions of the input sequence elements, potentially improving the accuracy of photo-$z$ estimation.

\section*{Acknowledgements}

ZJL acknowledges the support from the Shanghai Science and 
Technology Foundation Fund under grant No. 20070502400, and 
the science research grants from the China Manned Space Project. LPF acknowledges the support from the National Natural Science Foundation of China (NSFC 11933002). WD acknowledges the support from NSFC grant No. 11890691. YG acknowledges the support from National Key R\&D Program of China grant Nos. 2022YFF0503404, 2020SKA0110402, the CAS Project for Young Scientists in Basic Research (No. YSBR-092), and China Manned Space Project with Grant No. CMS- CSST-2021-B01. S.Z. acknowledges support from the National Natural Science Foundation of China (Grant No. NSFC-12173026), the Program for Professor of Special Appointment (Eastern Scholar) at Shanghai Institutions of Higher Learning and the Shuguang Program of Shanghai Education Development Foundation and Shanghai Municipal Education Commission. ZF acknowledges the support from NSFC grant No. U1931210. This work is also supported by the National Natural Science Foundation of China under Grants Nos. 12141302 and 11933002, and the science research grants from China Manned Space Project with Grand No. CMS-CSST-2021-A01. 

\section*{Data Availability}
 
The data that support the findings of this study are available from the corresponding author, upon reasonable request.



\bibliographystyle{mnras}
\bibliography{ref_lstm} 



\bsp	
\label{lastpage}
\end{document}